# Optical manipulation of sessile droplets of nematic liquid crystalline materials on the surface of a photovoltaic crystal

R. Karapinar[1], L. Cmok[2,*], X. Zhang[3], I. Drevenšek-Olenik[2,4]

1 Department of Nanoscience and Nanotechnology, Mehmet Akif Ersoy University, Burdur, Turkey
2 Department of Complex Matter, J. Stefan Institute, Jamova 39, Ljubljana, Slovenia
3 The MOE Key Laboratory of Weak-Light Nonlinear Photonics and International Sino-Slovenian Joint Research Center on Liquid Crystal Photonics, TEDA Institute of Applied Physics and School of Physics, Nankai University, Tianjin 300457, China
4 University of Ljubljana, Faculty of Mathematics and Physics, Jadranska 19, Ljubljana, Slovenia

**Abstract**. We investigated the effects of laser irradiation on sessile droplets of three well-known liquid crystalline materials (5CB, 8CB, E7) deposited on the surface of an iron-doped lithium niobate (LN:Fe) crystal. The static electric field, which is generated via the bulk photovoltaic effect in the LN:Fe substrate, produces the merging of smaller droplets into filaments oriented in the radial direction with respect to the laser spot. It also induces filament jetting from the rim of larger droplets toward the center of the illumination area. When the laser beam is focused directly onto the larger droplets, they abruptly disintegrate via the formation of several jet streams. The described effects are present in the nematic and also in the isotropic phase. We attribute them to a large gradient of the surface electric field that produces driving forces via the induced dipole moments of the droplets, analogous to electric field-based droplets transport mechanisms known for standard dielectric liquids.

*Second Author, E-mail: luka.cmok@ijs.si

## 1. Introduction

The synthesis of low molecular mass liquid crystalline (LC) materials exhibiting ferroelectric nematic phase ($N_F$)[1], which was for the first time successfully realized in 2017[2-4], stimulated extensive research of intriguing new properties and phenomena that are associated with the macroscopic polarization of this phase. One of them is explosive-type interfacial instability, which is observed when sessile droplets of the $N_F$ phase are exposed to high enough electric fields, i.e., to DC or AC fields with a magnitude above some threshold level[5-7]. Additionally, it was demonstrated that when $N_F$ droplets are deposited on a photovoltaic substrate, the fluid jets and/or smaller secondary droplets of $N_F$ material formed during the "explosions" can be manipulated by optical irradiation[8-11]. The main origin of these effects was proposed to be polar coupling between the $N_F$ phase and the external electric field. However, some other sources not limited to polar order, such as electrophoretic and dielectrophoretic forces, wettability gradients, etc., can produce similar effects. Therefore, to improve the fundamental understanding of the observed phenomena and elucidate different mechanisms that contribute to them, it is important to perform comparative investigations with droplets of other LC phases and the isotropic phase. Stimulated by our preliminary results, which showed that also droplets of a standard nematic LC (NLC) phase could become very "lively" when exposed to photovoltaic fields[6], we performed an extended investigation of optical illumination-based manipulation of sessile droplets of some well-known NLC materials deposited on the surface of photovoltaic lithium niobate crystals.

Lithium niobate ($LiNbO_3$ (LN), point group 3m) is a synthetic crystalline material that is widely used in various technological devices[12] due to its favorable ferroelectric, piezoelectric, pyroelectric, electro-optic, elasto-optic, and photorefractive properties. Iron-doped lithium niobate (LN:Fe), in addition, also exhibits a strong bulk photovoltaic effect[13-15]. Its irradiation with light of an appropriate wavelength causes the spatial redistribution of charge carriers in the illuminated volume. This process produces a static electric field in the range of 10-100 kV/cm in the direction along the polar axis of the crystal (c-axis)[16,17]. Before getting screened by the accumulation of surface charges, this space-charge field extends into the surrounding space around the crystal as an evanescent field[18]. Via electrophoretic and/or dielectrophoretic forces, this evanescent field can cause trapping and manipulation of small objects deposited on the crystal surface and induce their (re)distribution following the illumination pattern[19,20]; the effect that is known as photovoltaic tweezers (PVT).

Most research on the manipulation of sessile liquid droplets on the LN surface was performed with pure water and aqueous dispersions of different mostly organic materials. The results reported before 2018 are reviewed in the paper of García-Cabañes[21]. From papers published after 2018, we would like to point out the work of Puerto et al.[22] on droplet ejection and liquid jetting induced by laser irradiation of a LN:Fe platform, and the work of Zaltron et al.[23] on actuation and guiding of motion, merging and splitting of droplets by the LN:Fe-based optofluidic platform. In contrast to standard liquids, reports on experiments with droplets of standard (non-ferroelectric) NLC compounds are very rare. Bharath et al. investigated the influence of a uniformly poled Z-cut LN wafer on molecular anchoring of 4-cyano-4'-octylbiphenyl (8CB) molecules deposited on its surface by ambient vaporization[24]. They detected a significant difference between molecular orientation on the positive (c+) and the negative (c-) sides of the wafer, which was attributed to differences in interaction between the substrate and a relatively large molecular dipole moment of 8CB. Merola et al. studied the thermally-induced motion of NLC droplets of highly polar 1-(*trans*-4-hexylcyclohexyl)-4-isothiocyanatobenzene (6CHBT) molecules driven by the electric field generated via the pyroelectric effect of the LN[25-27]. They observed reversible droplet fragmentation and self-assembling that exhibited opposite behavior on the c+ and c- sides of the wafer too. The observed effects were attributed to the dielectrophoretic force. Kuktharev et al. used NLC droplets to analyze the quasi-periodic pulsation of a photovoltaic electric field generated in an LN:Fe crystal during optical illumination with a constant intensity[28]. For a green CW laser light, the pulsation had a period of around 5s. The effect was ascribed to microplasma discharges.

In the above-mentioned works, the researchers used either small NLC droplets (diameter < 100 µm) in combination with the pyroelectric field or relatively large (diameter ~5 mm) NLC drops in combination with the photovoltaic field. To the best of our knowledge, no reports exist on the effect of the photovoltaic field on small sessile NLC droplets. With this work, we would like to fill this gap by reporting on the optical illumination-induced rearrangement and merging/splitting of microsized droplets of some well-investigated commercial NLC compounds on the surface of LN:Fe substrates.

## 2. Experimental section

The congruent LN:Fe crystals with 0.05 mol% dopant concentration were grown by using the Czochralski method. The details of the growth procedure are reported elsewhere[29]. The as grown crystals were polarized, cut to 15.3 × 15.1 × 1.0 mm$^3$ slices (X × Y × Z; Z-cut plates) and optically polished. Before each set of experiments, the plates were washed in isopropanol, dried by the flow of nitrogen, heated to 200°C and then slowly cooled to the room temperature. They were used without any surface coating.

An LN:Fe plate was placed onto a supporting glass slide, and a drop of selected LC material with a volume of about 5 µl was deposited onto its surface. The following LC materials were used: E7 LC mixture from two different providers (Shijiazhuang Chengzhi Yonghua Display Material Co, and Merck Ltd., phase sequence: N ↔ 60°C ↔ Iso.), 4-cyano-4'-pentylbiphenyl (5CB, Nematel, GmbH & Co., phase sequence: Cryst. ↔ 22°C ↔ N ↔ 35.5°C ↔ Iso.), and 4-cyano-4'-octylbiphenyl (8CB, Merck, phase sequence: Cryst. ↔ 21°C ↔ SmA ↔ 33°C ↔ N ↔ 40°C ↔ Iso.). The drop was fragmented into several smaller droplets by gently wiping with the tattered edge of the wiping paper (Kimtech Science, Kimberly-Clark). For experiments associated with pyroelectric voltage or measurements associated with photovoltaic voltage at temperatures above room temperature, the assembly was put into the microscope heating stage (STC200, Instec Inc.) and heated to the desired temperature.

Before laser irradiation experiments, the assembly was left for a sufficiently long time at a fixed temperature to reach a stable state. In-situ optical irradiation was implemented by a CW laser beam operating at the wavelength of 532 nm (Compass 315M-100, Coherent) that was directed onto the assembly from above via the arm for the episcopic illumination in the polarization optical microscope (Optiphot-2-Pol, Nikon). The diameter (waist) of the Gaussian beam on the sample was ~0.8 mm, and its optical power was 40 mW, resulting in the optical intensity of ~10 W/cm2. The heating/cooling and the laser irradiation-induced modifications of the shape and positions of the LC droplets were monitored with polarization optical microscopy (POM) by using the diascopic illumination configuration. The unwanted reflected laser light was filtered away by an appropriate notch filter. The angle between the polarizer and the analyzer was set to 70°, as this enabled to resolve more details than in the case of crossed polarizers. The images and/or video-sequences were recorded by a standard color camera (Canon EOS M200).

## 3. Results

### 3.1. Formation of radial filament patterns

When the laser beam begins to irradiate an assembly with several NLC droplets, after some delay time, the droplets start to elongate into linear filaments. The filaments exhibit radial orientation with respect to the illuminated area and grow in both directions, towards the center and towards the external regions of the illuminated area. During this process, nearby droplets bridge with each other and tend to incorporate into the same filament. An example of this development at room temperature (25°C) for the droplets of E7 mixture sitting on the +c surface of an LN:Fe plate is shown in Figure 1. The orientation of the c+ and c- sides was determined by measuring the piezoelectric voltage[12]. The green circles indicate the location of the laser beam. The brown color is a consequence of the iron doping of the LN crystal. In the middle of the illuminated area, the filaments sometimes merge together, while in the far-out regions, they decompose into smaller secondary droplets.

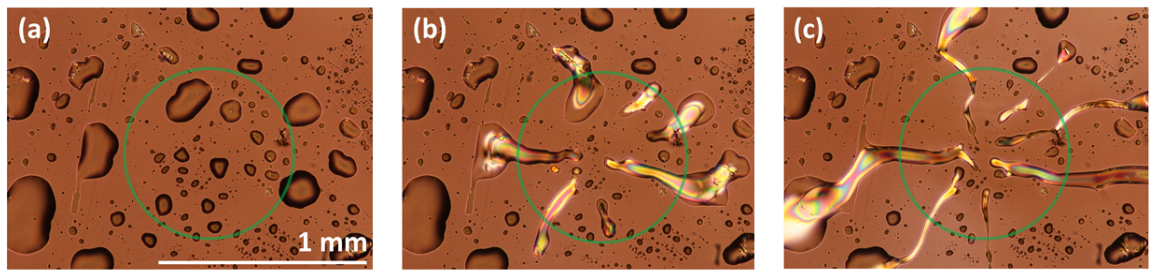

**Fig. 1** Illumination-induced rearrangement of an assembly of E7 droplets on the +c surface of a Z-cut LN plate at room temperature: (a) at the starting of laser irradiation, (b) after 7 s of irradiation, and (c) after 18 s of irradiation. The green circles indicate the location of a laser spot. (see also video S1)

As previous works on NLCs in combination with LN substrates reported some differences between the phenomena observed on the +c and -c sides of the LN plates[24-27], we repeated the experiment with E7 droplets deposited on the -c surface[12]. The result is shown in Figure 2. The initial internal orientation of LC inside the droplets is quite different from the case on the c+ side (see Figure 2a); however, their redistribution induced by laser irradiation seems to be very similar (see Figures 1c and 2b). Nearby droplets again merge into streams running in the radial direction. When the laser beam is switched off, the filament structure starts to disintegrate back into droplets. Only occasionally, some filaments remain pinned in an elongated form even a long time after the exposure (see Figure 2c).

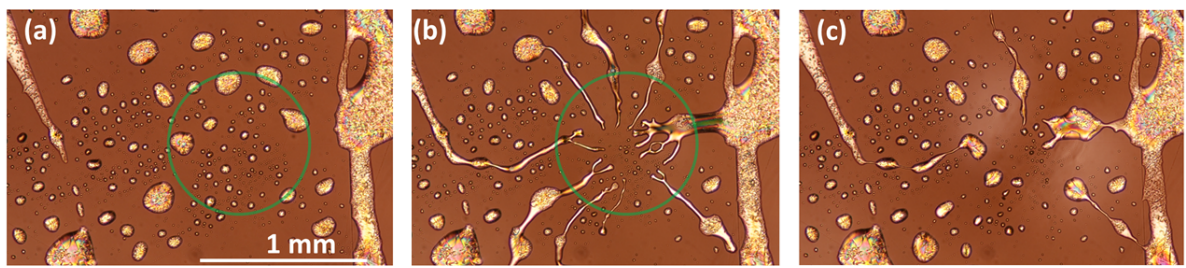

**Fig. 2** Illumination-induced rearrangement of an assembly of E7 droplets on the -c surface of a Z-cut LN plate at room temperature: (a) at the starting of laser irradiation, (b) 20 s after the irradiation was switched on, and (c) 35 s after the irradiation was switched off. The green circles indicate the location of a laser spot. (see also video S2).

By analyzing images obtained in ten separate experiments, we deduced information on the intermediate angle θ between two neighboring filaments. The result is shown in Figure 3. One can notice that the filaments show a preferential arrangement for 60° between two neighbouring filaments, in accordance with the 6-fold rotational symmetry of the Z-cut LN plates. The width of the peak is quite broad, which we attribute to the interplay between more or less randomly positioned droplets. Once a specific filament starts to grow, its growth direction is affected by the presence of nearby droplets that tend to join the filament.

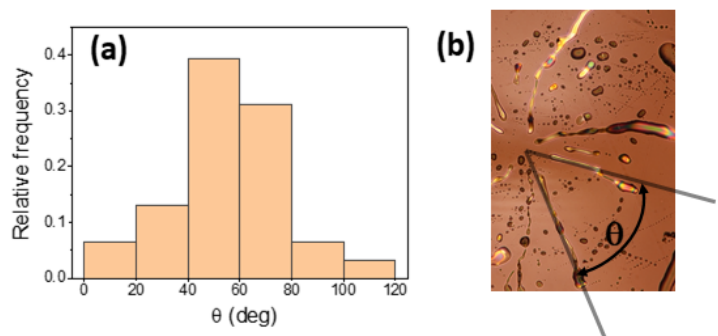

**Fig. 3 (a)** Statistical analysis of intermediate angle θ between the neighboring filaments of E7 at room temperature. (b) Example of determination of θ for two selected neighboring filaments.

The experiments with 5CB droplets at room temperature revealed the irradiation-induced formation of radially oriented filaments as well. However, the filaments were tinier and more transient in their appearance than those observed for E7, which is presumably due to the lower viscosity of 5CB with respect to E7. The results are shown in Figure 4. It can also be noticed that, similar to Fig. 2, the LC material is pushed out from the center of the illuminated area.

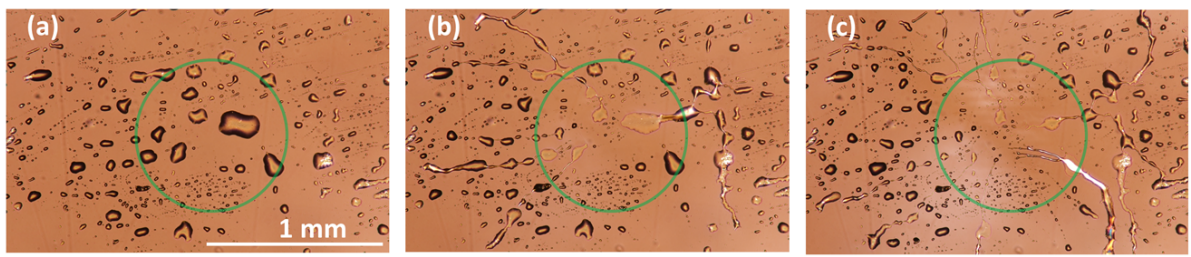

**Fig. 4** Illumination-induced rearrangement of an assembly of 5CB droplets on the +c surface of a Z-cut LN plate at room temperature: (a) at the starting of laser irradiation, (b) after 15 s of irradiation, and (c) after 30 s of irradiation. The green circles indicate the location of a laser spot. (see also video S3)

To obtain small, separated droplets of the nematic phase of 8CB the assembly was first heated to the isotropic phase (50°C, Iso) and then cooled down to the nematic phase (37°C, N). As shown in Figure 5, laser irradiation induces radial filament structure formation in both phases. Surprisingly, the filaments in the isotropic phase of 8CB (Figure 5b) are very similar to those observed in the nematic phase of 5CB and E7, while the filaments in the nematic phase of 8CB (Figure 5d) exhibit more unusual features. They are more numerous and very flat, i.e., their vertical thickness seems to be very low. The change in colors appeared due to the insertion of the heating stage. The black area on the left side of the images corresponds to the edge of the window of the heating stage. Subsequently, we also performed experiments in the isotropic phase of E7 and 5CB and observed illumination-induced filament structures too. This finding reveals that the formation of filament structures is not limited to the liquid crystalline state of the droplets.

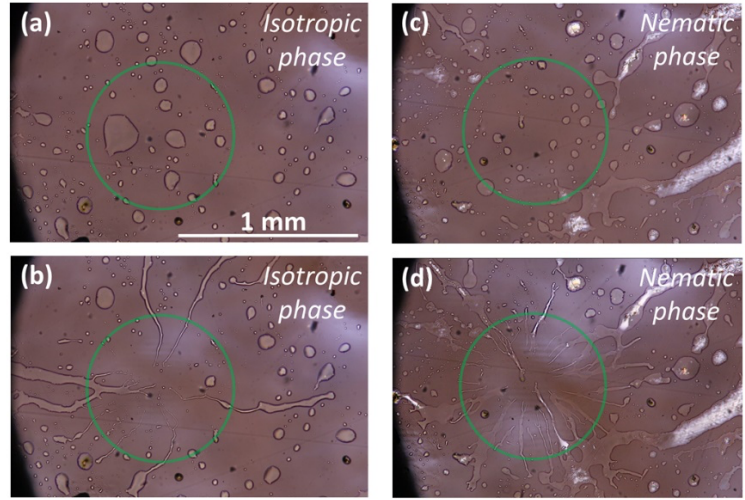

**Fig. 5** Illumination-induced rearrangement of an assembly of 8CB droplets on the +c surface of a Z-cut LN plate at 50°C (isotropic phase) and at 37°C (nematic phase): (a) at the starting of laser irradiation, (b) after 25 s of irradiation; (c) at the starting of laser irradiation, (d) after 25 s of irradiation. The green circles indicate the location of a laser spot. (see also video S4 and S5)

### 3.2. Drawing of filaments from larger droplets

We also investigated what happens if the laser beam is focused on the region of the LN:Fe substrate in the vicinity of a large LC droplet. By and large, we mean droplets with a surface area similar to the size of the laser spot. An example of an E7 droplet at room temperature (25°C) is shown in Figure 6. After some delay time concerning the starting of illumination, the internal orientational structure (nematic director field) of the droplet starts to fluctuate, which can be noticed as significant modifications of the colors inside the droplet (Fig. 6b). Shortly after this, a conical jet of the LC material is ejected in the direction towards the center of the illuminated area (Fig. 6c). After this, if the laser beam is moved further away from the rim of the droplet, the jet elongates into a filament (fiber) and follows the laser beam until it splits into smaller filaments (Fig. 6d). These secondary filaments then start to interact with nearby smaller droplets and at this point, the process of fiber "drawing" stops, and the previously described process of multiple-filament formation follows. We repeated the experiments also with 5CB droplets and observed very similar phenomena.

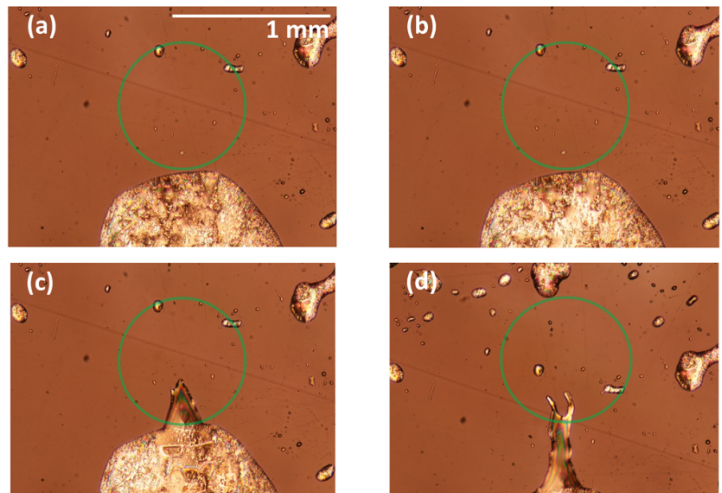

**Fig. 6** Illumination-induced formation of E7 filament on the +c surface of a Z-cut LN plate at room temperature: (a) at the starting of laser irradiation, (b) after 4 s of irradiation; (c) after 7 s of irradiation, (d) after 20 s of irradiation. Between c) and d) the sample was translated downwards with respect to the laser beam. The green circles indicate the location of a laser spot. (see also video S6)
9...

*3.3. Illumination-induced disintegration of larger droplets*

In the last series of experiments, we studied the response of large droplets to a laser beam that was focused on the inner area of the droplet. Fig. 7 shows the result obtained with a 5CB droplet at room temperature (25°C). The illumination, at first, causes strong modifications and fluctuations of the orientational structure (nematic director field) inside the droplet (Fig. 7a). Subsequently, in very rapid (explosive) events, several jets are ejected from the droplet in different directions. After a longer time, most of the LC material is removed from the initial area. The latter effect can be nicely seen in Figure 8, which shows the images of a disintegration process for the E7 droplet.

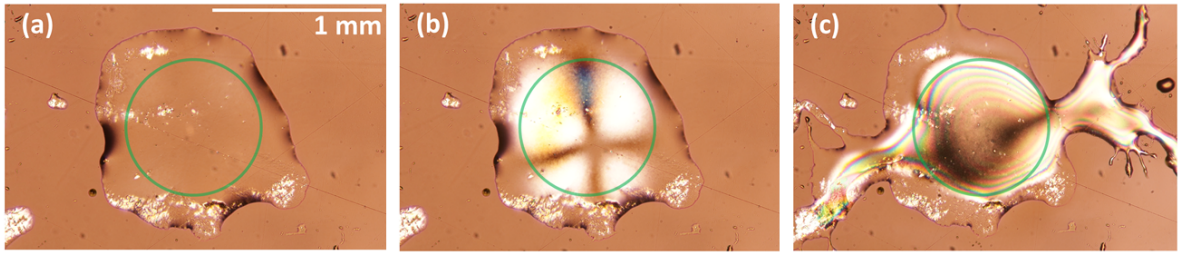

**Fig. 7** Illumination-induced disintegration of 5CB droplet on the +c surface of a Z-cut LN plate at room temperature: (a) at the starting of laser irradiation, (b) after 5 s of irradiation; (c) after 37 s of irradiation. The green circles indicate the location of a laser spot. (see also video S7)

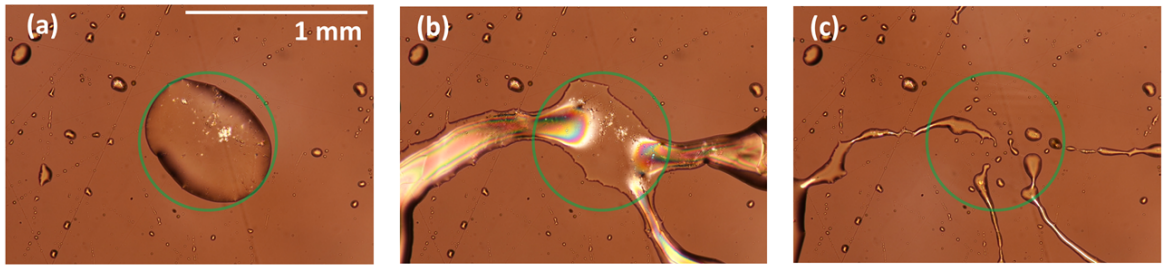

**Fig. 8** Illumination-induced disintegration of E7 droplet on the +c surface of a Z-cut LN plate at room temperature: (a) at the starting of laser irradiation, (b) after 15 s of irradiation; (c) after 70 s of irradiation. The green circles indicate the location of a laser spot. (see also video S8)

Similar effects were also observed at higher temperatures in the nematic as well as in the isotropic phase. In all cases, for droplets that are in size comparable to or smaller than the laser spot, prolonged exposure leads to the removal of the droplet from the illuminated area. Figures 9.a,b show the results obtained for the E7 droplet slightly below the N-I phase transition temperature (60°C), and figures 9.c,d for the 8CB droplet in the isotropic phase (50°C). The details of the disintegration and removal processes are quite variable and depend on the specific LC material and phase as well as on the particulars of the shape and size of the initial droplet.

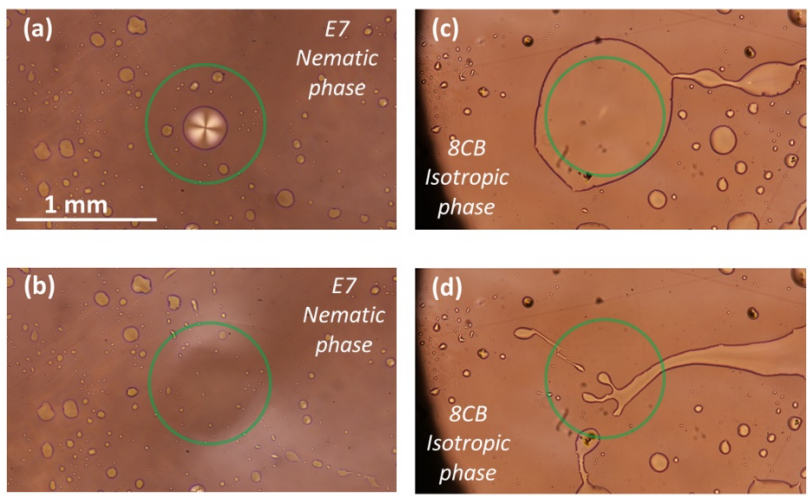

**Fig. 9** Illumination-induced disintegration of E7 (at 60°C) and 8CB (at 50°C) droplets on the +c surface of a Z-cut LN plate: (a) at the starting of laser irradiation, (b) after 15 s of irradiation; (c) at the starting of laser irradiation, (d) after 33 s of irradiation. The green circles indicate the location of a laser spot.

## 4. Discussion and conclusions

As mentioned in "Experimental", additionally to the experiments described above, for all the investigated compounds we also analyzed the effect of pyroelectric voltage. In this purpose, the droplets were deposited on the LN:Fe surface, and the temperature of the entire assembly was rapidly modified (~5°C/s), which produced a pyroelectric field of a similar magnitude to the photovoltaic field generated by the laser irradiation (~100 kV/cm)[6]. Nevertheless, in these experiments, we only occasionally noticed some dynamic features; in particular, we observed filament-type bridging of neighboring droplets in the edge regions of the optical window of the heating stage. This is the region in which temperature gradients in the sample created during heating/cooling cycles are the largest; therefore, the pyroelectric field there is supposed to be very inhomogeneous. This finding suggests that gradients of surface electric fields play a vital role in the droplet manipulation processes reported in this work. At this point, we would like to stress out that in this respect, there seems to exist a big difference between the standard nematic (N) and the ferroelectric $N_F$ phase, for which explosive dynamic properties and formation of jet streams were also observed in pyroelectrically generated electric fields[5,6,8].

Conversely to conventional heating/cooling of the samples obtained by the heating stage, their optical irradiation with the Gaussian beam causes the generation of a strongly inhomogeneous electric field. This field has a dipolar form and is localized to the illuminated areaP0. The corresponding dielectrophoretic force acting on dielectric objects located in and around the illuminated region is oriented predominantly along the c-axis of the crystal[30]. On the Z-cut plate, the vertical component of this force is compensated by gravity, so the dynamic processes are governed by the in-plane component of the field that exhibits a radial direction with respect to the center of the illuminated area. We assume that this component and its gradients determine the preferential orientation of the jets and filaments observed in our experiments.

Similar effects were widely investigated with droplets of conventional liquids, such as water, exposed to strong electric fields and are known as Taylor cones, electrohydrodynamic tip-streaming, cone-jetting, Rayleigh jets, etc.[22,31-35] Granda et al. have recently demonstrated that subcritical electric fields do not only stretch the liquid drops into filaments but can also cause their fingering into trilobite-like configurations[36], very similar to the one shown in Fig. 7c. All these intriguing effects are attributed to the interplay between the (di)electrophoretic forces and surface tension of the liquid, which investigation is technologically important, among others, in the field of digital microfluidics for droplet transport and routing[37].

Another effect that should not be neglected in the case of LN:Fe substrates, is the thermal gradient generated due to a relatively strong absorption of the laser light by the iron ions. Via temperature dependence of surface tension, such gradients typically produce transport of droplets from the hotter towards colder areas, which in our case means out of the illuminated area[38], in agreement with the observations. Nevertheless, there exist also many other possible origins of this type of droplet motion, for instance, inhomogeneous electro-wetting or inhomogeneous charge accumulation[37], so the observed phenomena are very probably a combination of many of them.

The question that naturally arises from our study is how nematic liquid crystalline order influences the above-described features. The main differences are expected to emerge from the strong anisotropy of various physical properties of the nematic phase. Nevertheless, based on the results reported in this work, in our opinion, for the standard NLC phase it is not possible to claim any profound features that would be specific for the liquid crystalline state. There for sure exist differences in some selected details, but these are out of the scope of the present work; as for their investigation, one should establish much more controlled conditions, in particular as regards the size, shape, and spatial arrangement of the droplets on the LN:Fe substrate. On the other hand, our results demonstrate that optical anisotropy of the NLCs is very convenient for visualization of the investigated phenomena. For instance, strong intradroplet fluctuations appearing before jet ejection were readily resolved by the polarization optical microscopy, while much more effort would be needed to detect such fluctuations in the usual isotropic liquid. We hence believe that further investigations of the phenomena, such as those described in this study, can bring new insights into the optical illumination-induced electrohydrodynamic instabilities of liquid droplets on the surface of photovoltaic substrates.

**Acknowledgments**

We acknowledge the financial support of the Slovenian Research Agency (ARRS, P1-0192), and National Natural Science Foundation of China (12074201), PCSIRT (IRT 13R29), 111 Project (B23045). R.K. would like to acknowledge the financial support of the COST Action EUTOPIA (CA17139)